\documentclass[aps,twocolumn,10pt,superscriptaddress,prl,floatfix]{revtex4-1}

\usepackage{bm}
\usepackage{graphicx}
\usepackage{subfigure}
\usepackage{siunitx}
\graphicspath{{figures/}}

\def\be{\begin{equation}}
\def\ee{\end{equation}}
\def\e#1{\label{#1}\end{equation}}
\def\bea{\begin{eqnarray}}
\def\eea{\end{eqnarray}}
\def\ea#1{\label{#1}\end{eqnarray}}

\def\bem#1{\begin{mathletters}\label{#1}}
\def\eml{\end{mathletters}}

\def\4#1{{\boldsymbol{#1}}}
\def\8#1{{\widetilde{#1}}}
\def\bse{\begin{subequations}}
\def\ese{\end{subequations}}

\newcommand{\simleq}{\; \raisebox{-0.4ex}{\tiny$\stackrel
{{\textstyle<}}{\sim}$}\;}

\begin{document}

\title{Spectroscopy of Surface-Induced Noise Using Shallow Spins in Diamond}

\author{Y. Romach}
\affiliation{The Racah Institute of Physics, The Center for Nanoscience and Nanotechnology, The Hebrew University of Jerusalem, Jerusalem 91904, Israel}
\author{C. M\"uller}
\author{T. Unden}
\author{L. J. Rogers}
\affiliation{Institute for Quantum Optics and Center for Integrated Quantum Science and Technology, University of Ulm, D-89081 Ulm, Germany}
\author{T. Isoda}
\author{K. M. Itoh}
\affiliation{School of Fundamental Science and Technology, Keio
University, Yokohama 223-8522 Japan}
\author{M. Markham}
\author{A. Stacey}
\affiliation{Element Six, Ltd, Kings Ride Park, Ascot SL5 8BP, United Kingdom}
\author{J. Meijer}
\author{S. Pezzagna}
\affiliation{Institute for Experimental Physics II, Linn{\'e}stra{\ss}e 5, University of Leipzig, 04103 Leipzig, Germany}
\author{B. Naydenov}
\author{L. P. McGuinness}
\email{liam.mcguinness@uni-ulm.de}
\thanks{Corresponding author.}
\affiliation{Institute for Quantum Optics and Center for Integrated Quantum Science and Technology, University of Ulm, D-89081 Ulm, Germany}
\author{N. Bar-Gill}
\email{bargill@phys.huji.ac.il}
\thanks{Corresponding author.}
\affiliation{Department of Applied Physics, Rachel and Selim School of Engineering, Hebrew University, Jerusalem 91904, Israel}
\affiliation{The Racah Institute of Physics, The Center for Nanoscience and Nanotechnology, The Hebrew University of Jerusalem, Jerusalem 91904, Israel}
\author{F. Jelezko}
\affiliation{Institute for Quantum Optics and Center for Integrated Quantum Science and Technology, University of Ulm, D-89081 Ulm, Germany}

\begin{abstract}
We report on the noise spectrum experienced by few nanometer deep nitrogen-vacancy centers in diamond as a function of depth, surface coating, magnetic field and temperature. Analysis reveals a double-Lorentzian noise spectrum consistent with a surface electronic spin bath in the low frequency regime, along with a faster noise source attributed to surface-modified phononic coupling. These results shed new light on the mechanisms responsible for surface noise affecting shallow spins at semiconductor interfaces, and suggests possible directions for further studies. We demonstrate dynamical decoupling from the surface noise, paving the way to applications ranging from nanoscale NMR to quantum networks.
\end{abstract}

\maketitle

Nanoscale magnetic imaging and magnetic resonance spectroscopy, recently demonstrated using nitrogen-vacancy (NV) color centers in diamond \cite{Staudacher2013,Mamin2013,Grinolds2013,rondin13}, are capable of yielding unique insights into chemistry, biology and physical sciences. The sensitivity and resolution of these techniques relies heavily on the NV coherence properties, which empirically are much worse for shallow NV centers than those deep within bulk diamond \cite{wrachtrup_ultralong_2009}. An understanding of the origin of surface related noise enables optimal decoupling or surface passivation to be performed. It is critical not only for improving NV applications in quantum sensing \cite{taylor2008,wrachtrup_efield}, quantum information processing \cite{laddquantum2010}, and photonics \cite{faraon11}, but is also an outstanding problem in many solid-state quantum systems (e.g. \cite{Martinis_noise1,Martinis_noise2}). Furthermore, overcoming noise at the diamond interface is a significant obstacle to realizing hybrid quantum systems with NV centers \cite{zhu11, kubo11}, which are expected to play an important role in realistic devices.

\begin{figure}[tbh]
\begin{center}
\includegraphics[width=0.9 \columnwidth]{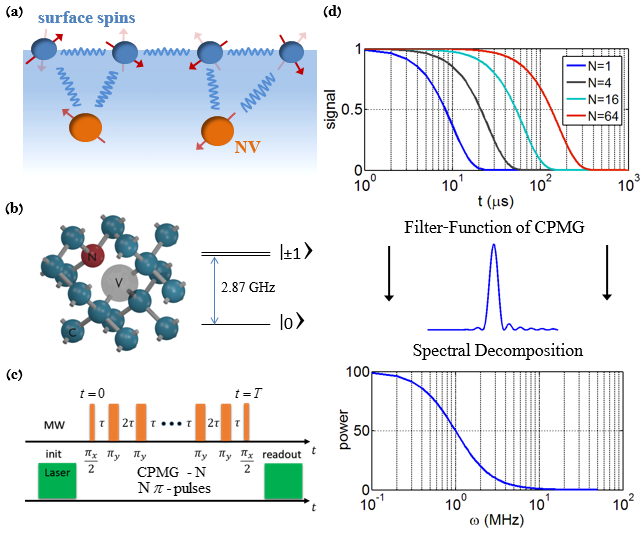}
\protect\caption{(a) Model system comprising shallow NV color centers in diamond (spin, S = 1), interacting with fluctuating electronic spins on the diamond surface. (b) Structure of the NV center. (c) Measurement scheme: initialization and readout using 532\,nm light, between which CPMG dynamical decoupling control sequences are applied. (d) Spectral decomposition technique: The NV coherence is measured with CPMG sequences of varying pulse number and pulse spacing. The environmental noise power spectrum is then obtained from coherence measurements.} \label{fig:intro}
	\end{center}
\end{figure}

For NV centers in bulk diamond, noise sources limiting coherence times have been identified with internal nuclear and electronic spin baths, and interactions with phonons \cite{bar-gill_spect_decomp, Bar-Gill2013}. Although additional noise sources related to the diamond surface, and affecting shallow NVs, have been observed \cite{Degen2012}, their origin is not currently well understood. This phenomenon is general and has been observed at various semiconductor interfaces, resulting in the development of several theoretical models, which are still without significant experimental confirmation \cite{lee14, desousa07}. Here we use shallow implanted NV centers as nanoscale sensors to perform spectroscopy of the diamond surface. We use dynamical decoupling techniques together with measurements of longitudinal ($T_1$) relaxation under varying conditions (surface coating, magnetic field, temperature) in order to characterize the surface-induced noise. The strength and frequency dependence of fluctuations as a function of the NV distance from the surface are investigated with nanometer precision. We directly measure the noise spectrum experienced by shallow NV centers, revealing an unexpected double-Lorentzian structure which indicates contributions from two distinct noise sources. We find that the low frequency noise experienced by shallow NVs is consistent with electronic spin impurities on the surface [Fig.\,\ref{fig:intro}(a)], with a relaxation mechanism consistent with dipolar coupling between the spins. The NVs also experience high frequency noise components (attributed to surface-modified phonons), which contribute to both decoherence and relaxation of the NV sensor.

The understanding gained from this work allows decoupling of NV centers from environmental noise, enabling higher sensitivity to be achieved. Moreover, we expect similar noise sources and spectral behavior to be relevant to a wide range of other systems, including quantum dots, superconducting qubits, and phosphorus in silicon architectures. Finally, the spectral decomposition technique we employ for noise analysis is general, and could be utilized to extend our physical understanding of noise dynamics in such systems.

The NV center consists of a substitutional nitrogen atom and a vacancy occupying adjacent lattice sites in the diamond crystal. The electronic ground state is a spin triplet, in which the $m_\mathrm{s}=0$ and $\pm 1$ sublevels experience a $\sim 2.87$ GHz zero-field splitting [Fig.\,\ref{fig:intro}(b)], while a static magnetic field can further split the $\pm 1$ sublevels to create an effective two-level system.The NV spin can be initialized with optical excitation, detected via state-dependent fluorescence intensity, and coherently manipulated using microwaves \cite{childress2006}.

We performed measurements on NV centers in ultrahigh purity diamonds, created with low energy (2.5\,keV) implantation so that their nominal depth from the surface is 2--5\,nm, calculated using SRIM \cite{SRIM}. The actual depths of the NVs were later precisely measured by detecting the proton NMR signal from immersion oil on the diamond surface \cite{SuppMatPRL14}\nocite{cpmg,universal_decay}. The ability to extract meaningful information from experiments depends critically on the sample conditions. In particular, the diamond substrate contains very few spin impurities ($^{13}$C $< 10^{-3}\%$, N $< 5$ ppb), and the low implantation dose of $10^8$ N$^{+}$ ions/cm$^2$ created NVs with a concentration on the order of $10^7 [\rm{cm}^{-2}]$. The shallow NV depth and high substrate purity produces samples in which surface noise dominates, allowing a straightforward and unambiguous analysis.  Our data in this work are based on 10 NV centers measured in two diamond samples created using the above technique. We note that the technique described here can also be used to investigate the noise spectrum of NVs incorporated near the diamond surface during the final stages of diamond growth via ``delta-doping'' \cite{Rosskopf2013,Myers2014}. ``Delta-doping'' consists of controlled introduction of nitrogen during the diamond-growing process, creating a thin (few nm) nitrogen-doped (and NV rich) layer \cite{DeltaDoping}.

Our analysis of the noise experienced by the NV is based on spectral decomposition \cite{bar-gill_spect_decomp}, in which the NV coherence is measured as a function of time, while applying periodic dynamical decoupling Carr-Purcell-Meiboom-Gill (CPMG) pulse sequences [Fig.\,\ref{fig:intro}(c)]. The NV coherence decays as a function of time due to interactions with a noisy environment, characterized by a noise spectrum $S (\omega)$. Spectral decomposition recovers $S(\omega)$ from the NV decoherence curves [Fig.\,\ref{fig:intro}(d)] \cite{SuppMatPRL14}.

In addition, longitudinal spin relaxation measurements, in which no control pulses are applied, were used to determine the $T_1$ time scale. These measurements are sensitive to high frequency noise at the NV Larmor frequency, which cannot be probed using the coherent spectral decomposition approach.

\begin{figure}
\includegraphics[width=0.9 \columnwidth]{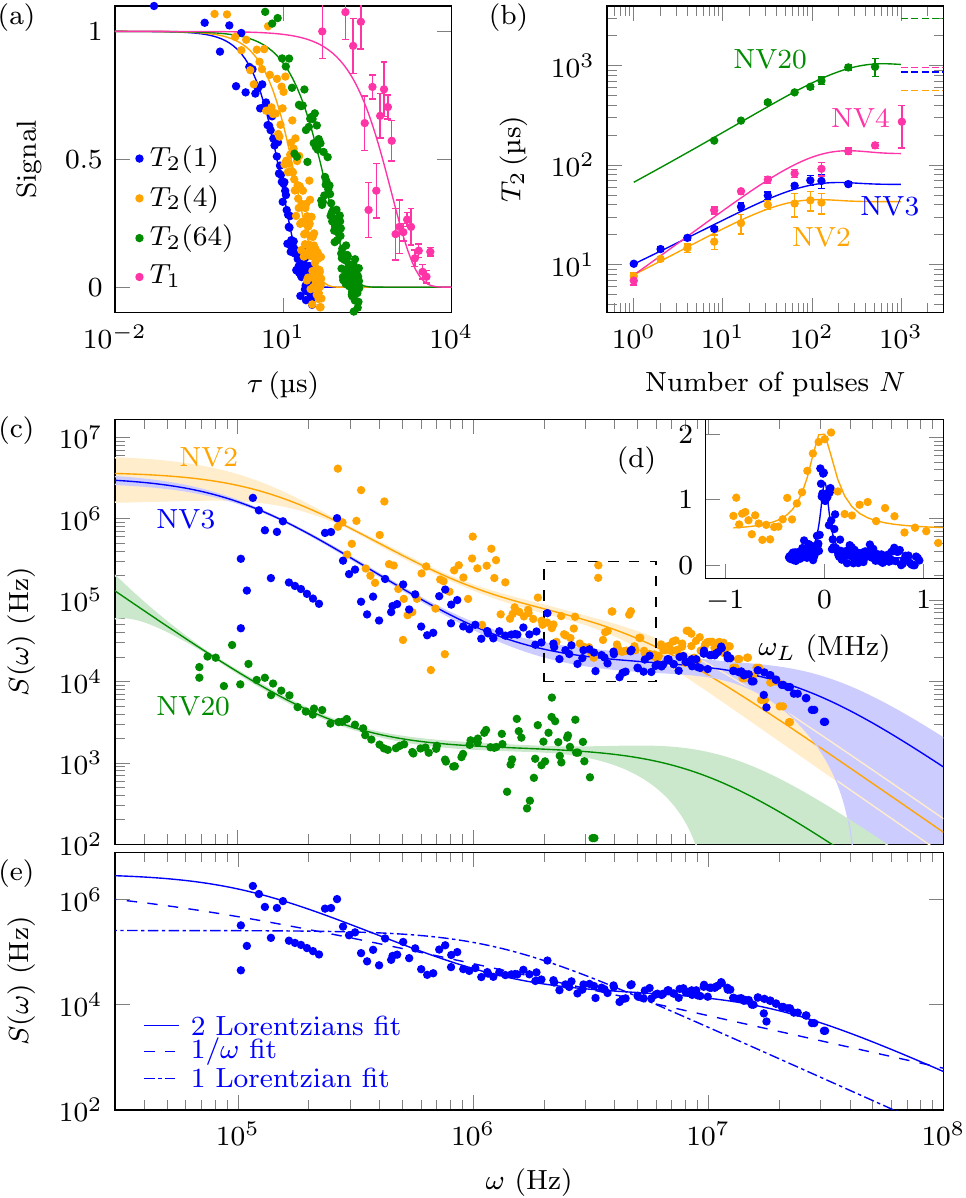}
\protect\caption{Measured decoherence curves, extracted coherence times, and noise spectra for several NVs at various depths.
	(a) Coherence vs time of NV3 for different number of pulses $N$. $T_1$ error bars indicate standard errors, other error bars are negligible on this scale. 
	(b) Coherence time as a function of the number of pulses, $T_2(N)$ extracted from decoherence curves. Solid lines are fits to saturation curves (see \cite{SuppMatPRL14}). Measured $T_1$ is indicated with dashed lines. 
	(c) Noise spectra extracted from decoherence curves using spectral decomposition (see text). The fits are to Eq. (\ref{eq:S}), colored regions indicate $1\sigma$ confidence regions (error bars can be found in \cite{SuppMatPRL14}).
	(d) A zoom in of spectral noise observed at hydrogen Larmor frequency (\textit{x} axis is relative to that frequency, \textit{y} axis is arb. unit).
	(e) Fits of the noise spectrum from NV3 to single Lorentzian ($\chi^2 = 26.581$), $1/\omega$ ($\chi^2 = 12.799$) and double Lorentzian ($\chi^2=0.969$) curves. See \cite{SuppMatPRL14} for further analysis.} \label{fig:s69}
\end{figure}

In Fig.\,\ref{fig:s69} we show measurements of four shallow NVs, labeled according to their approximated depth in nanometers: NV2, NV3, NV4, and NV20, performed at room temperature in a static magnetic field of 454 G. The measured coherence as a function of time for NV3 is plotted in Fig.\,\ref{fig:s69}(a), with each curve depicting a pulse sequence of different pulse number $N$. The data are fitted with $\exp\left[-\left(\frac{t}{T_2}\right)^p\right]$ with $p$ values ranging between $1-3$ \cite{bar-gill_spect_decomp}, from which we extract the coherence time for a given number of CPMG pulses $N$, denoted as $T_2(N)$, as shown in Fig.\,\ref{fig:s69}(b). Moreover, the coherence data are deconvolved with the filter function associated with each applied pulse sequence to extract the spectrum of noise experienced by each NV [Fig.\,\ref{fig:s69}(c)].

Appearing in the spectrum is a signal occurring at the hydrogen Larmor frequency [measured using the XY8 sequence \cite{Gullion1990,Ryan2010}, Fig.\,\ref{fig:s69}(d)]. This signal provides an absolute method for subnanometer determination of the NV depth \cite{SuppMatPRL14} (used throughout this work), and demonstrates the performance of the NV center as a nanoscale sensor \cite{Staudacher2013,Mamin2013,ohashi13}.

From Fig.\,\ref{fig:s69}(b) we see that shallower NVs experience stronger noise and have much shorter coherence times, providing evidence that the signal indeed originates from the surface. NV2, NV3 and NV4, which are measured to be 2--4\,nm deep, exhibit one-pulse coherence times of $T_2(1) \sim$~5--\SI{10}{\micro \second}, whereas NV20, which is 20(5) nm deep, exhibits a much longer coherence time of $T_2(1) = \SI{64(20)}{\micro \second}$. We observe that NV20 exhibits a scaling of $T_2$ with the number of pulses $T_2(N)=N^k$ of $k=0.53(6)$ (slightly lower than the expected limit for a simple Lorentzian spin bath, which is $k=2/3$) \cite{sousa2009}, while the scaling for the shallower NVs is significantly lower at $k \simeq 0.3-0.48$ [Fig.\,\ref{fig:s69}(b)] (exact extracted numbers in \cite{SuppMatPRL14}). The measured spin relaxation times for NVs 2, 3, and 4 are $T_1^\mathrm{NV2} = \SI{430(225)}{\micro \second}$, $T_1^\mathrm{NV3} = \SI{860(200)}{\micro \second}$, and $T_1^\mathrm{NV4} = \SI{960(500)}{\micro \second}$, while for NV20 it is $T_1^\mathrm{NV20} = 3(1)$ ms.

The different noise experienced by shallower NV centers is also evidenced in the saturation values of the coherence time $T_2^\mathrm{sat}$, which, together with $T_1$, are related to high frequency noise that cannot be suppressed by the applied pulse sequences (Fig. \ref{fig:s69}[b] and \cite{SuppMatPRL14}). For NV20 coherence saturation occurs at $T_2^\mathrm{sat} = 0.9(3)$\,ms, which is slightly lower than the ratio $T_2^\mathrm{sat} \simeq 0.5 T_1$ observed for NVs deep in bulk diamond \cite{Bar-Gill2013}. In contrast, coherence saturation for the shallower NVs is close to $1/10 T_1$ [$T_2^\mathrm{sat} = \SI{42(12)}{\micro \second}$ and $\SI{64(10)}{\micro \second}$, for NV2 and NV3, respectively]. Shallow NVs experience stronger high frequency noise compared to bulk NVs, which couples both to relaxation processes (i.e., $T_1$) and to decoherence processes (resulting in smaller $T_2^\mathrm{sat}/T_1$ ratio), suggesting a different noise source compared to that in bulk diamond.

To understand the origin of the noise more precisely we compare the observed spectra to models of different physical processes [Fig.\,\ref{fig:s69}(c)]. In particular we compare to $1/\omega$ noise, which is observed in superconducting circuits \cite{oliver}, to the Lorentzian noise characteristic of spin baths \cite{klauder_anderson,sousa2009}, and telegraphic noise observed in quantum dots \cite{marcus_dd_scaling}. 
From a reduced chi-squared analysis, we find that fits to these common spectral functions show manifestly poor agreement with our data [$\chi^2 = 12.799, 26.581$ for Lorentzian and $1/\omega$ spectra, respectively, Fig.\,\ref{fig:s69}(e)]. Therefore, we extend the analysis by allowing two independent noise sources, which better capture the behavior of the measured spectra ($\chi^2 = 0.969$, see \cite{SuppMatPRL14}). The two noise sources are modeled as Lorentzian functions:
\begin{equation}
S (\omega) = \sum_{i=1,2} \frac{\Delta_i^2 \tau_\mathrm{c(i)}}{\pi} \frac{1}{1 + (\omega \tau_\mathrm{c(i)})^2}, \label{eq:S}
\end{equation}
where $\Delta_i$ is the average coupling strength of the environment to the NV spin, and $\tau_\mathrm{c(i)}$ is the correlation time of the environment.
We initially attribute the slower correlation time $\tau_\mathrm{c(1)} \simeq$ 10--\SI{20}{\micro \second} to spin-spin coupling between bath spins, and the faster correlation time $\tau_\mathrm{c(2)} \simeq 100-250$\,ns to surface-modified phonons coupled to the NV spin.

\begin{figure}
\includegraphics[width=0.9 \columnwidth]{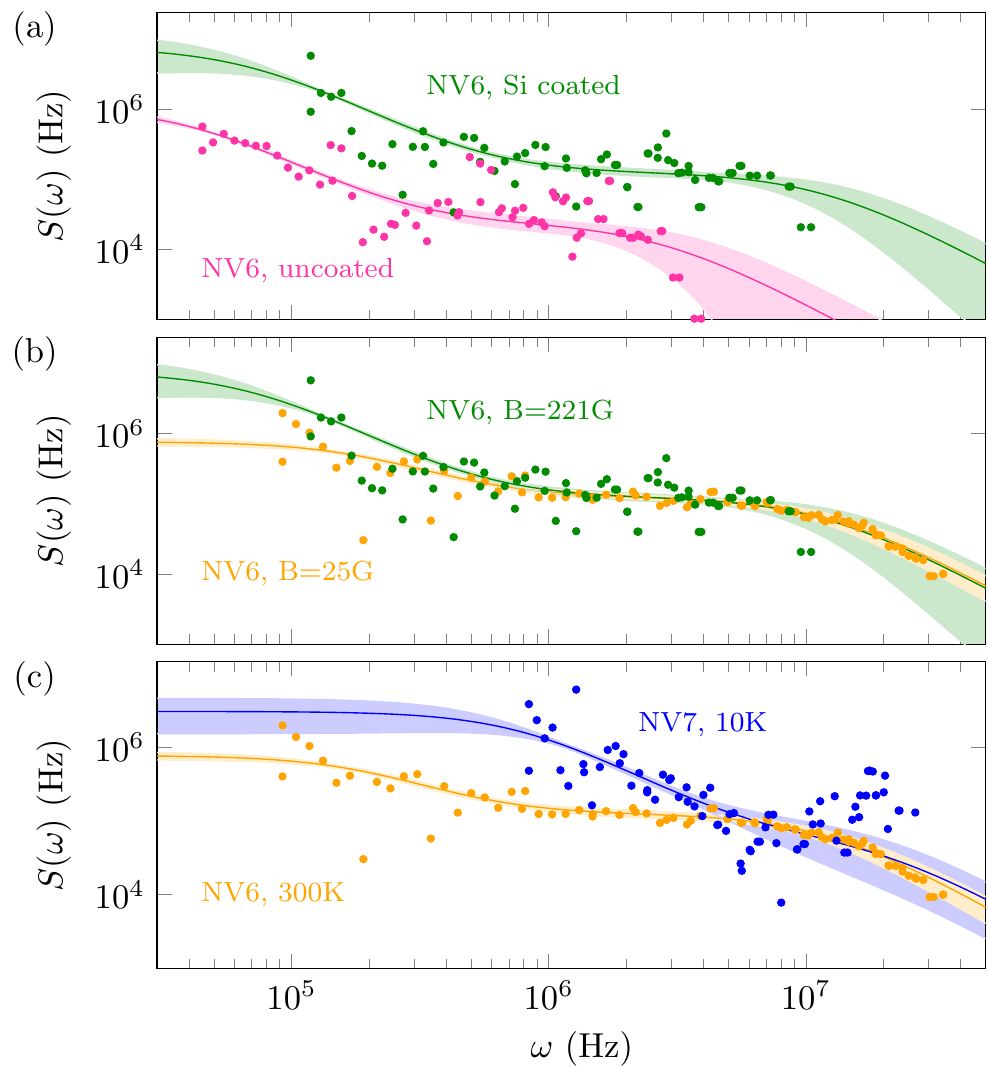}
\protect\caption{
Comparison of noise spectra for varying parameters; colored regions indicate $1\sigma$ confidence regions.
(a) Noise spectra for NV6, uncoated and coated with silicon.
(b) Noise spectra for NV6 in different static magnetic fields.
(c) Noise spectra for NV6 and NV7 (both coated with Si) at room temperature and at 10 K.
}
\label{fig:s70}
\end{figure}

In order to gain further information about the NV's local environment, we conducted a series of experiments varying external parameters. To rule out diffusion of spins in the immersion oil or a water layer adsorbed on the diamond surface as responsible for surface noise, we deposited a 4\,nm thick layer of silicon on the diamond with molecular beam epitaxy (the top $\sim 2$ nm oxidized to SiO$_2$). 
In Fig.\,\ref{fig:s70}(a) we compare the noise spectrum measured with and without the silicon layer on the diamond surface. In general, the measured NVs exhibited larger noise with the Si coating, but overall similar behavior \cite{SuppMatPRL14},
implying that the noise is intrinsic to the diamond surface. The fact that the noise strength is increased but the dynamics remain largely unaltered with the silicon layer also lends credence to the hypothesis that an electronic spin bath is responsible, since this mechanism has also been proposed at Si/SiO$_2$ interfaces \cite{morello10}.

In Fig.\,\ref{fig:s70}(b) we plot the noise spectrum measured at low and high magnetic fields for NV 6. The change in magnetic field is expected to affect $T_1$ behavior by varying the resonance frequency of the NV transition, and potentially $T_2(N)$ behavior, by varying the spin bath dynamics through detuning of spin species with different energy scales (e.g. hyperfine energies of N impurities).
We do not observe a statistically significant change in $T_2$ or $T_1$ times \cite{SuppMatPRL14}. 
We therefore conclude that the spin bath dynamics do not depend on the \textit{B} field, at least up to the values investigated here. 

Figure\,\ref{fig:s70}(c) presents data measured on another NV at cryogenic temperatures (10 K). The effect of temperature on the surface-induced noise was studied in order to gain insight into the role of spin-phonon coupling at the diamond surface, which is strongly temperature dependent \cite{reynhardt98,budkerT1}. Higher frequency noise is indeed greatly reduced (except for a peak at $\sim 20$ MHz), and the coherence time is extended to $T_2(32) = \SI{23(6)}{\micro \second}$, with no indication of saturation (see \cite{SuppMatPRL14}). We also observed a significantly longer spin relaxation time $T_1 \gg 1$ ms at 10 K (compared to $T_1 \simleq 0.5$ ms of other shallow NVs at room temperature). These results are consistent with the expectation that the high frequency noise responsible for $T_1$ and $T_2^\mathrm{sat}$ is strongly dependent on temperature, and suggests phononic effects.

\begin{figure}
\includegraphics[width=0.9 \columnwidth]{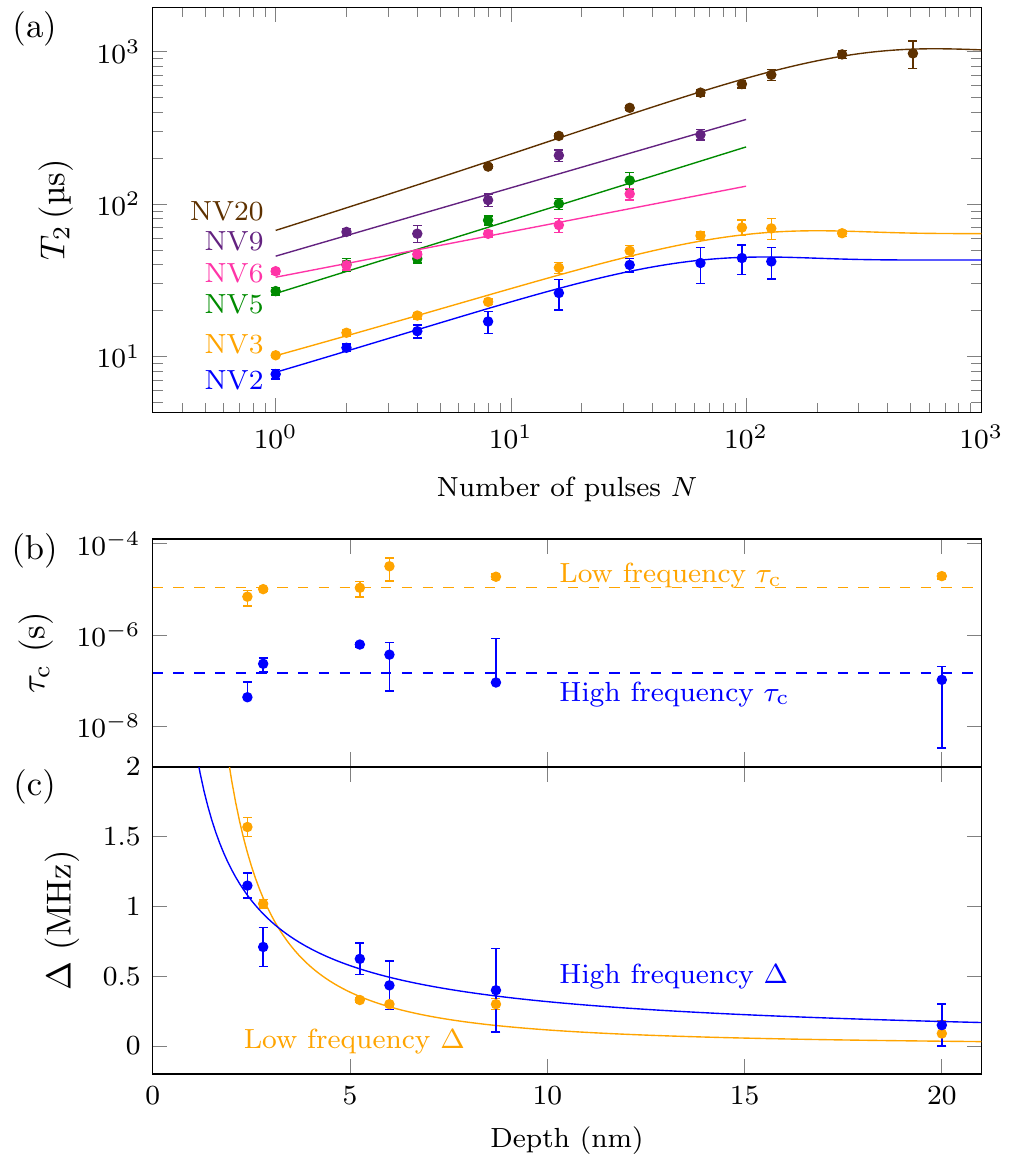}
\protect\caption{(a) Scaling of $T_2$ with the number of pulses $N$ for different NV depths (full data sets in \cite{SuppMatPRL14}). (b) Extracted low and high frequency bath correlation times as a function of NV depth. Dashed lines are values extracted from global fitting $\tau_\mathrm{c(1,2)}= \SI{11(1)}{\micro \second},146(14)\,\textrm{ns}$. (c) Extracted low and high frequency bath coupling strengths, with global fitting, as a function of NV depth. Solid lines are fits to $\frac{a}{d^n}$.
}
\label{fig:depth}
\end{figure}

In Fig.\,\ref{fig:depth} we plot the coherence data and extracted environmental parameters as a function of NV depth $d$. We note that the correlation times for both low frequency and high frequency noise components are largely independent of depth, as expected of a parameter internal to the noise source [Fig. \ref{fig:depth}(b)]. Therefore, we refitted the data using global fitting with shared correlation times \cite{SuppMatPRL14,Spitzer}. The extracted coupling strengths are plotted vs depth in Fig.\,\ref{fig:depth}(c), depicting inverse scaling with depth for both the low and high frequency noise. Fitting this depth scaling to $\frac{a}{{d}^n}$, the low frequency noise exhibits a power law of $n=1.75(21)$, consistent with $1/{d}^2$ as expected from a 2D electronic spin bath \cite{Myers2014}. However, the power law behavior of the high frequency component is $n=0.9(3)$. This scaling suggests a different physical mechanism for this noise component, possibly surface-modified phonons, and is inconsistent with fast phonon-induced dynamics in the 2D spin bath \cite{Rosskopf2013,mcguinness13}. 
The low frequency correlation time $\tau_\mathrm{c(1)} \sim \SI{11(1)}{\micro \second}$, assuming a 2D electronic spin bath, corresponds to an average spin spacing of 2--3 nm (based on the dipolar coupling strength of electronic spins, $g=2$). 

We note that for NV depths comparable to or below the surface spin density, the approximation of a uniform bath of spins breaks down, and a slightly different depth scaling than $1/d^2$ is expected due to the small number of spins that interact with the NV. 
Nevertheless, to first order our results capture the dominant features of the NV environment.

We briefly compare our results to other recent work with shallow NV centers that we have become aware of during the preparation of the manuscript \cite{Rosskopf2013,Myers2014}. The NV relaxation rates and environmental correlation times observed here are consistent with one study attributing the surface noise to an electronic spin bath, albeit with reduced spin density \cite{Rosskopf2013}. The correlation time and depth scaling we observe is also in agreement with the observations of Ref.\,\cite{Myers2014}, whereas coupling to phonons has also previously been implicated \cite{mcguinness13}. In particular, the reconstructed double-Lorentzian noise spectra that we obtain here provide direct evidence of a combination of an electronic spin bath and a phonon-related relaxation mechanism. We also note one double electron-electron resonance study which confirmed the presence of $g=2$ electron spins on the diamond surface \cite{grinolds14}. 

In conclusion, we have studied the surface-induced noise affecting shallow NV centers in diamond. Through controlled experiments varying surface coating, magnetic field and temperature, along with detailed noise spectrum analysis, we conclude that the surface noise is consistent with an electronic spin bath that undergoes slow spin-spin dynamics, along with another fast phonon-induced noise that is coupled to the NV directly. The exact nature of the noise, which we attribute to surface-modified phonons, remains an open question, and further studies are required to rule out, for example, the role of electric fields to NV decoherence. We investigated the possibility of suppressing surface-induced noise through coating of the diamond surface with a silicon layer, but no improvement in coherence times was observed. Further studies using the methodology we have demonstrated here can potentially be used to design tailored surface terminations to enhance shallow NVs coherence.

The frequency dependence of $S(\omega)$ that we observe here means that even for very shallow NV centers dynamical decoupling is effective at suppressing environmental decoherence, allowing record coherence times of $T_2 \sim \SI{50}{\micro \second}$ for NV centers two nanometers from the surface. The noise spectrum discovered here could guide the tailoring of better decoupling methods to improve coherence times even further. The sensitivity we achieve for such shallow NVs is of importance in quantum information, metrology, and photonics applications
, and, in particular studying spin dynamics on the diamond surface at the single spin level.

This research has been supported by DARPA/AFOSR QuASAR, EU (CIG, EQUAM, ERC Synergy Grant BioQ, DIAMANT), and DFG (SFB TR 21, FOR 1493, FOR 1482, SPP1601) Grants, the Minerva ARCHES Award, and the Alexander von Humboldt Foundation. The work at Keio has been supported in part by a Grant-in-Aid for Scientific Research by MEXT, by Cannon Foundation, and by the JSPS Core-to-Core Program. We gratefully acknowledge helpful discussions with Linh M. Pham, Christian Degen, and Jared Cole.

Y. R., C. M., and T. U. contributed equally to this work.

\bibliography{NV}

\end{document}